\documentstyle[graphics,12pt]{article}
\begin{document}

ADP-95-14/T175  \hspace*{6cm} March 1995

KFA-IKP(TH)-1995-04

\vspace{2cm}
\centerline{\large Mesonic Corrections to the Shape of
Quark Distributions}
\vspace{.4cm}
\centerline{F.M. Steffens$^a$, H. Holtmann$^{a,b}$ and A.W. Thomas$^a$}
\vspace{.2cm}
\centerline{$a$) Department of Physics and Mathematical Physics}
\centerline{University of Adelaide}
\centerline{Adelaide, S.A. 5005, Australia}
\centerline{$b$) Institut f\"ur Kernphysik, Forschungszentrum J\"{u}lich GmbH.}
\centerline{52425 J\"{u}lich, Germany}

\begin{abstract}

We compute the full x - dependence of the proton and neutron
spin structure functions in the MIT bag model, including the effect of
gluon exchange and the meson cloud. Impressive agreement
is found for x larger than 0.1, where polarised gluons are
not expected to play a significant role.

\end{abstract}
\newpage

At the present moment  there is
an impressive collection of high energy data on nucleon structure functions
that defy a complete understanding. To quote a
few, we mention the NMC \cite{nmc} measurement of the Gottfried
sum rule, the NA51 \cite{na51} measurement of the sea
asymmetry and the (already classical) problem with the
polarized structure functions as measured by EMC/SLAC [3 - 6].
In the particular case of the
polarized structure functions,
it now seems that one has reached the end of the ``spin crisis"
and the beginning of the spin problem \cite{slac143,tony1}.
As it is unlikely that perturbative QCD corrections alone are
enough to solve the problem, we should examine the possibility
that at least some of the discrepancy arises in the non-perturbative regime.
A common approach has been to calculate the desired quantities
at some low scale and then evolve them, using the QCD renormalization
group equations, to the experimental scale. Unless the model has been
specifically derived from QCD, the starting scale at which it best approximates
the
structure of the nucleon is unknown. It is usually adjusted to fit the
experimental
parton distributions as well as possible. Of course, one might expect
that some models would give a better description of the data than others
and there is considerable interest in using DIS to help choose amongst models
\cite{tony2}.

Although there are
a variety of possibilities to calculate quark distibutions at
the starting scale,  like non-relativistic quark models \cite{nrq}
or vertex functions \cite{vertex}, here we shall work with the MIT bag.
This choice is based on its success in low energy physics, its
simplicity and the insights it has already yielded in connecting
low and high energy data \cite{andreas,fernando}. Of course we
do not expect that the bag model alone contains all the physics we need
because we know that mesons should play a role as well.
For instance, we know \cite{tony3} that chiral symmetry in
the bag is restored through a meson cloud. Moreover, the meson cloud
is already known to give important corrections to some sum rules [14 - 19]

As important as it is to correct the integrated quark distributions,
the study of the effect of the meson cloud on the $x$ dependence of the
quark distributions themselves
is at least as important [21 - 24].
In this letter we calculate the $x$ dependence of the polarized and
unpolarized quark distributions
in the proton in a bag model dressed by mesons. We perform the calculations
in next-to-leading order (NLO) QCD. Moreover, we present
for the first time the $x$ dependence of the combined effect of the $N$ -
$\Delta$
interference term \cite{tony6} and one gluon exchange \cite{close}.
The interference term is important because, as noticed before \cite{tony4},
in the Ellis-Jaffe sum rule it cancels part of the reduction coming from pions.
As we shall see, the interference terms have a determinant role in the
shape of the polarized quark distribution of the neutron.

We introduce mesons in the model through the Sullivan
process \cite{sullivan}, simply noting that there are
unresolved questions about the model, particularly the validity
of the impulse approximation \cite{jaffe}. The modern study of the mesonic
contribution via
convolution started with Thomas \cite{tony10} and was later
extended to the study of structure functions by the Adelaide
group \cite{tony5,wally,wallyrev}. A problem remaining in these
calculations is the freedom for the value of the cut-off in the
form factor. To avoid this problem, we shall follow
the approach of the J\"{u}lich group \cite{harald2,harald1}, where
the cut-off is fixed through the use of high energy $pp$ data.

The basic hypothesis in this sort of model is that the physical nucleon
wave function (in the infinite momentum frame) can be written
as a superposition of a few Fock states:

\begin{equation}
|N\rangle_{phys}=Z^{1/2}\left[ |N\rangle_{bare}+
\sum_{BM} \int_0^1dx\int d^2k_\perp\phi_{BM}(x,k_\perp)
  |B(x,k_\perp),M(1-x,-k_\perp)\rangle \right].
\label{fock}
\end{equation}
The wave function renormalization factor
\begin{equation}
Z=\left[1+\sum_{BM}\int_0^1dx\int d^2k_\perp|\phi_{BM}(x,k_\perp)|^2
\right]^{-1},
\end{equation}
measures the probability that the physical nucleon contains a bare
nucleon. The Fock states used in our calculation involve the low mass
particles which are important to describe nucleon
properties, namely the nucleon $SU(3)$ octet ($N$, $\Lambda$, $\Sigma$)
and decuplet ($\Delta$, $\Sigma^*$) and the first pseudoscalar and
vector meson octet ($\pi$, $K$, $\rho$, $\omega$, $K^*$). We have
included the hyperon-kaon contributions for completeness, but
their actual size is very small ($\sim2\%$) \cite{harald2,signal2,wallyrev}.

The Fock state expansion Eq.(\ref{fock}) has consequences for the
structure function of the nucleon. Due to the presence of baryon-meson
Fock states, the virtual photon can scatter either on the nucleon core
or on the meson-baryon system.
Formally, the quark distribution $q(x)$ of the nucleon is given by:

\begin{equation}
q(x)=Z\left[q_{bare}^N(x)+\sum_{BM}
 \left(\delta^M q(x)+\delta^B q(x)\right)\right].
\label{f2}
\end{equation}
The contributions from the virtual meson and baryon can be written as
the convolution of the meson (baryon) structure function with its
longitudinal momentum distribution in the nucleon:

\begin{equation}
\delta^M q(x)=\int_x^1 f_{MB}(y) q^M\left({x\over y}\right)\frac{dy}{y}
\qquad{\rm and}\qquad
\delta^B q(x)=\int_x^1 f_{BM}(y) q^B\left({x\over y}\right)\frac{dy}{y},
\label{f3}
\end{equation}
where $f_{MB}$ and $f_{BM}$ are given by:

\begin{equation}
f_{MB}(x)=\int d^2k_\perp |\phi_{BM}(1-x,-k_\perp)|^2,
\qquad{\rm and}\qquad
f_{BM}(x)=\int d^2k_\perp |\phi_{BM}(x,k_\perp)|^2.
\end{equation}
In order to conserve charge and momentum, we have the following relation:

\begin{equation}
f_{BM}(1-y)=f_{MB}(y).
\end{equation}
The functions
$f_{BM}(y)$ and $f_{MB}(y)$ can be calculated using time ordered perturbation
theory in the infinite momentum frame \cite{wallyrev,harald1,zoller}.  The
analytic forms
for $f_{BM}(y)$ and $f_{MB}(y)$, can be found in Refs. \cite{wallyrev,harald1}.

In practical calculations we need more information, namely the various
coupling constants and the vertex form factors $G_{BM}$.
The coupling constants can be extracted from scattering
experiments and are rather well known \cite{pp}.
For the vertex form factor we use an exponential parametrization:

\begin{equation}
G_{BM}=exp\left[{1\over 2\Lambda_{BM}^2}(m_N^2-M_{BM}^2(y,k^2_\perp))\right],
\end{equation}
with $M_{MB}^2 (y,k^2_\perp) = \frac{m_B^2 + k^2_\perp}{y} + \frac{m_M^2 +
k^2_\perp}{1 - y}$.
The cut-off parameters $\Lambda_{BM}$ can be estimated, using one
boson exchange models, from $n$,
$\Lambda$ and $\Delta^{++}$ production in high energy $pp$ scattering
(for details see Ref. \cite{harald2}). They were found to be $\Lambda_{N\pi}
= \Lambda_{N\rho} = 1.08 \; GeV$ and $\Lambda_{\Delta\pi}=
\Lambda_{\Delta\rho}= 0.98 \; GeV$. The procedure used to obtain
the cut-off parameters may be questioned because the assumption
of single meson exchange being responsible for the process is fairly
simple. However, it certainly gives an upper bound for the values of
the various $\Lambda^{'} s$.

At this point we are still left without the input distributions in
equations (\ref{f2}) and (\ref{f3}).
For the quark distribution in the pion we use a recent parameterization by
Sutton et al. \cite{sutton}. By using $SU(3)$ symmetry, the quark distributions
of
all the mesons can then be obtained. For the bare quark distribution of the
baryons we use the bag model calculation of the Adelaide group
\cite{andreas,signal}. The main advantage of this method is that
it ensures energy-momentum conservation and hence the correct
support of the quark distributions. It has also been successful
in describing data at high momentum transfer, $Q^2$, both in leading
order \cite{andreas} and in next-to-leading order \cite{fernando}.
Mesonic corrections lead to further improvement in the predictions
of the model and will, among other things, reduce the amount
of evolution needed \cite{thomas}. The form of the quark distribution
is given by \cite{andreas}:

\begin{eqnarray}
q^{\uparrow \downarrow}_f (x)&=&\frac{M}{(2\pi)^2}\sum_m
\langle\mu|P_{f,m}|\mu\rangle
\nonumber \\*
     &\times&\int_{[M^2 (1-x)^2 -M_{n}^{2}]/2M(1-x)}^{+\infty}\mid\vec{p}_n\mid
     d\mid\vec{p}_n\mid \frac{|\phi_2(\vec{p}_n)|^2}{|\phi_3(0)|^2}|
     \tilde\psi^{\uparrow \downarrow}_{m}(\vec{p}_n)|^2 .
\label{f10}
\end{eqnarray}
Here $|\mu\rangle$ is the spin-flavor part of the wave function of the initial
state (at rest), $P_{f,m}$ makes the projection onto flavor $f$ and spin
projection
$m$, $M_n$ is the mass of the intermediate state and $\tilde\psi$ the Fourier
transform of
the quark wavefunction. Equation (\ref{f10}) gives the two quark contribution
to the total
quark distribution which dominates at intermediate and large $x$.
We should also account for the contributions coming
from four quarks in the intermediate state, but for simplicity we shall mimic
this
contribution by a term of the form $(1-x)^7$, properly normalized
(further discussion on this subject can be found in Ref.
\cite{andreas}).

\begin{figure}[h]
\vspace{5cm}
\graphics{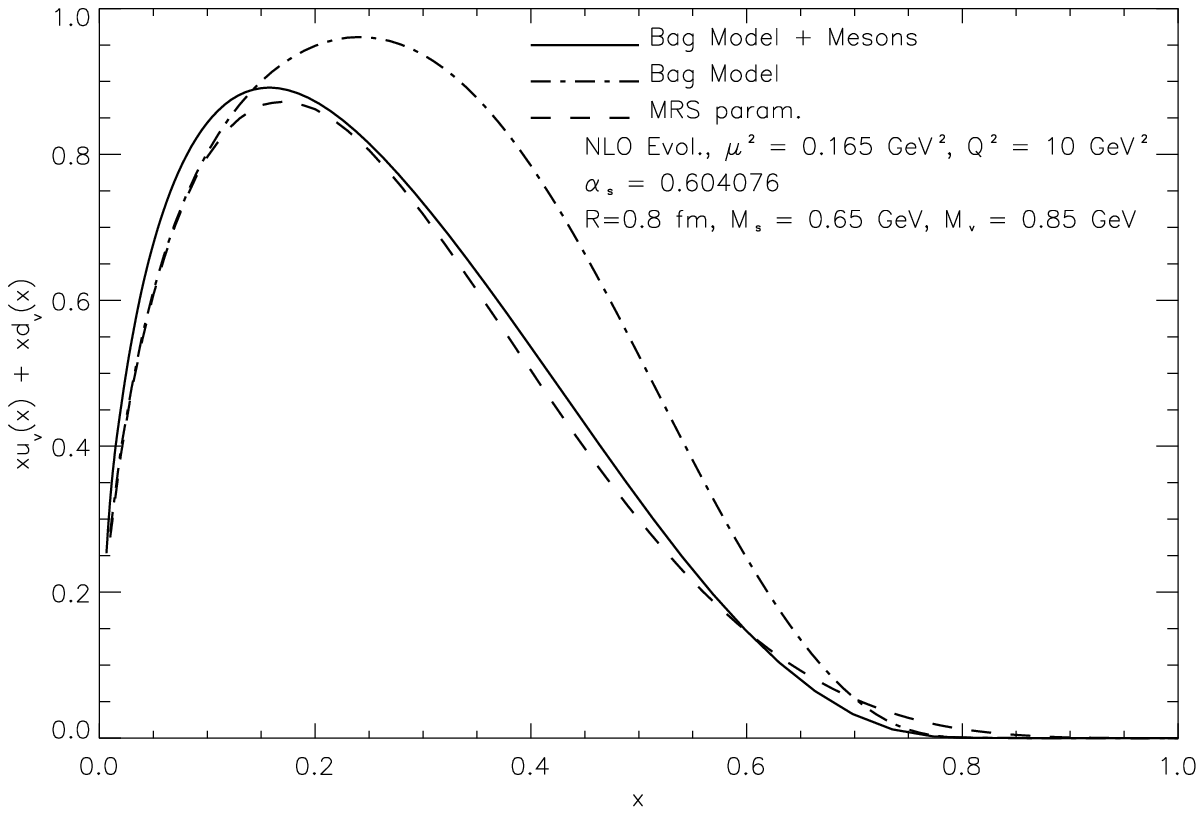}{17}{10}{0}{-10}{1}
\caption{Total valence distribution in the bag and in the bag dressed with
mesons compared with the MRS \protect\cite{mrs} parametrization of the data in
the $\overline{MS}$ scheme. The quark distributions are evolved
in nest-to-leading-order $QCD$.}
\label{fig2}
\end{figure}

We use the MRS parametrization \cite{mrs}
of the unpolarized structure functions to fix the
parameters of the model; e.g., the radius of the bag, the average mass of
the spin scalar ($M_s$) and spin vector ($M_v$) diquark in the intermediate
state
and the low scale, $\mu^2$, at which the model is supposed to be valid.
In Fig. \ref{fig2} we show the total valence distribution computed in
$NLO$\footnote{Details for $NLO$ calculation in the bag can be found
in \cite{fernando}} in the $\overline{MS}$ scheme for the bag
dressed with mesons. Very good agreement with the MRS parametrization
is found for $\mu^2 = 0.165\; GeV^2$, $R=0.8 \; fm$, $M_s = 0.65 \; GeV$
and $M_v = 0.85 \; GeV$. For comparison, we also show the bare bag
calculated with the same set of parameters. The bonus of the
NLO fit is that it provides a decrease of $\alpha_s$ at $\mu^2$ from
$\sim 1.52$ in a LO fit to $\sim 0.6$ in NLO. This is also a drop in comparison
with the value $\alpha_s \sim 0.77$
found in a NLO fit without mesons \cite{fernando}. For completeness, we note
that these values of the coupling constant were found using $\Lambda_{QCD} =0.2
\; GeV$ and
three active flavors.

After we have fixed all parameters in the unpolarized deep
inelastic scattering sector, we can explore the consequences for
the polarized sector. The calculation of effects due to the
presence of higher $BM$ Fock states for the $g_1(x)$ structure function
is similar to those for $q(x)$ given in
Eq.(\ref{f2}). The contribution from the scattering on the
recoil baryon is
\begin{equation}
\delta^B \Delta q(x) = \sum_M \int_x^1 d_{BM}(y)\Delta q^B\left({x\over
y}\right) {dy\over y},
\label{f11}
\end{equation}
where $d_B(y)$ is the polarized, longitudinal momentum distribution. It
can be calculated using the same techniques as for the unpolarized case
(for details see \cite{harald2}). The mesonic contribution vanishes because
of the pseudoscalar character of the pion.
The main difference from the unpolarized structure function is the
presence of the $N$-$\Delta$ interference term \cite{tony6},
which can also be written as a convolution
\cite{harald1}:
\begin{equation}
\delta^{int}\Delta q (x)=\int_x^1 d_{int}(y) \Delta q^{N\Delta}\left({x\over
y}\right) {dy\over y}.
\label{f12}
\end{equation}
The necessary polarized splitting functions $d_{BM}$ and
$d_{int}$ can be found in the Appendix.

Combining all these contributions, $g_1$ for the nucleon is given by:
\begin{equation}
g_1^{phys}=Z\left[g_1^{bare}+\sum_{BM}
\left(\delta^B g_1^B +\delta g_1^{int}\right)\right].
\label{f13}
\end{equation}

We now apply Eqs. (\ref{f11}) - (\ref{f13})
to calculate $g_1 (x)$.
In the meson sector only the pseudoscalars are included.
The vector mesons are omitted because we are not aware of
any model to extract their polarized quark distribution functions.
Once again we need a model for the bare quark distributions
in the nucleon and, as before, we use the MIT bag model.
For the polarized case we need to specify the spin-flavor
part of the wave function. We shall use the usual $SU(6)$ wave function
in which case, for the interference terms, we have:

\begin{equation}
\langle p^{\uparrow}, n^{\uparrow}| u^{\uparrow}|\Delta^{+,\uparrow},
\Delta^{0,\uparrow}\rangle =
\langle p^{\uparrow}, n^{\uparrow}| d^{\downarrow}|\Delta^{+,\uparrow},
\Delta^{0,\uparrow}\rangle = \frac{\sqrt{2}}{3},
\label{f14}
\end{equation}
\begin{equation}
\langle p^{\uparrow}, n^{\uparrow}| u^{\downarrow}|\Delta^{+,\uparrow},
\Delta^{0,\uparrow}\rangle =
\langle p^{\uparrow}, n^{\uparrow}| d^{\uparrow}|\Delta^{+,\uparrow},
\Delta^{0,\uparrow}\rangle = - \frac{\sqrt{2}}{3}.
\label{f15}
\end{equation}
We notice that in the matrix elements (\ref{f14}) and (\ref{f15})
only mixed symmetric terms contribute and, as a consequence, the intermediate
state always forms a spin vector.

The first moment of the polarized structure function for the proton,
$g_1^p (x)$, is expressed in NLO as:

\begin{equation}
\int_0^1 g_1^p (x,Q^2) dx = \left( \frac{g_a}{12} + \frac{g_8}{36}\right)
\left(1 - \frac{\alpha_s (Q^2)}{\pi}\right) +
\frac{g_0}{9}\left(1 - \frac{\alpha_s (Q^2)}{3 \pi}\right),
\label{f16}
\end{equation}
with $g_a$ and $g_8$ nonsinglet distributions and
$g_0$ a singlet distribution. However, the full singlet anomalous
dimensions (for any moment) for polarized scattering in NLO are still not known
and, because of that, it is not possible to calculate the
$x$ dependence of $g_0$ in NLO. Faced with this problem, we decided to
take the following two approaches to the evolution of $g_1$:

$\bullet$ In the first approach, {\bf case (a)}, we evolve $g_a$ and $g_8$ as
nonsinglet in NLO and evolve $g_0$ as a singlet in LO so that
$g_0$ does not pick up the $(1 - \frac{\alpha_s (Q^2)}{3 \pi})$
correction and the whole structure function is overestimated.

$\bullet$ In the second approach, {\bf case (b)}, we treat $g_a$, $g_8$ and
$g_0$ as nonsinglet combinations
and evolve them in NLO. In this case, $g_0$ picks up a
correction of the form $(1 - \frac{\alpha_s (Q^2)}{\pi})$
such that the corrections due to the NLO evolution are overstimated
and the structure function is underestimated.
The actual curve must be somewhere between the two approaches.

\begin{figure}[h]
\graphics{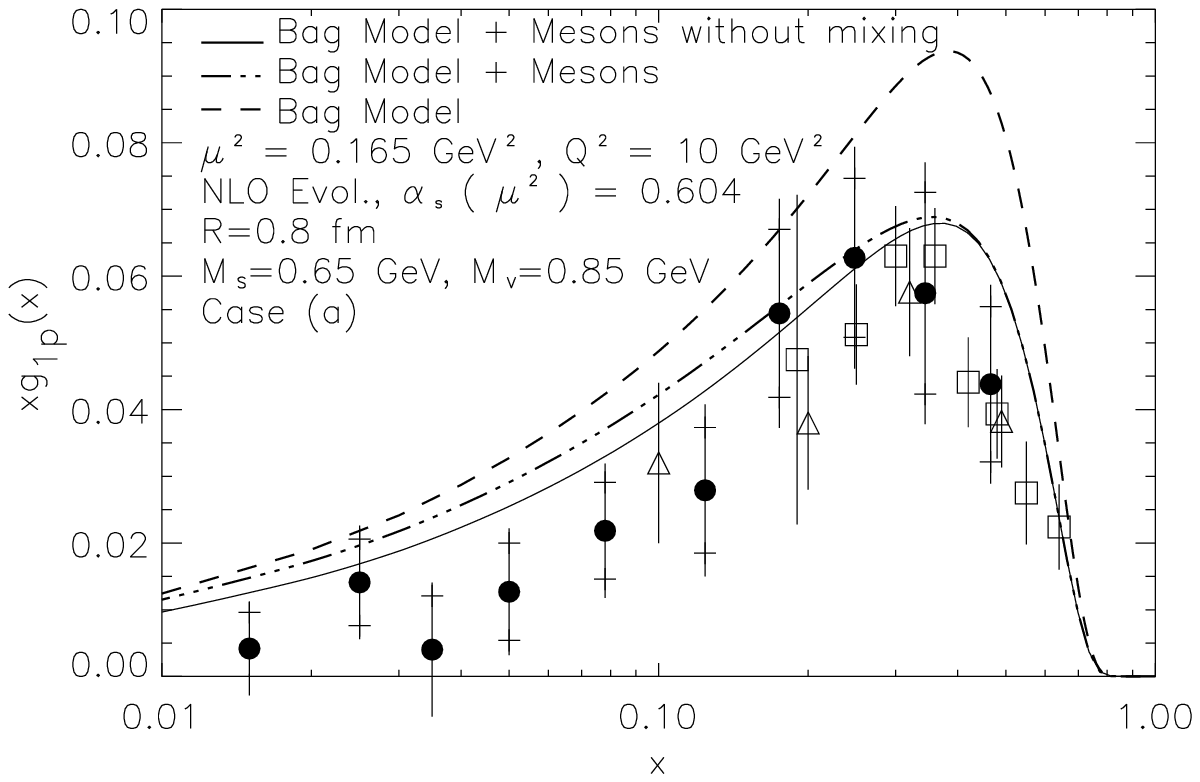}{17}{10}{0}{-10}{1}
\graphics{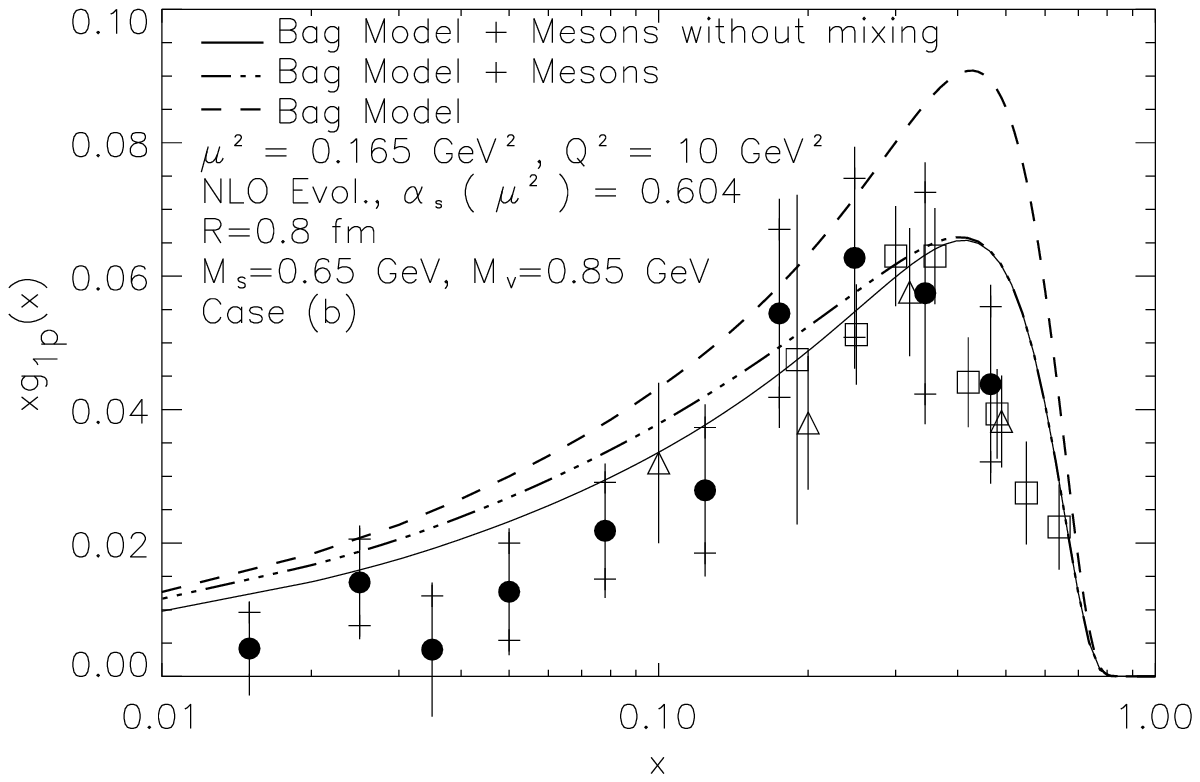}{17}{10}{0}{-10}{1}
\caption{Polarized quark distribution of the proton as
measured by the EMC collaboration \protect\cite{emc} against theoretical
predictions for a bare bag, a bag with mesons without
mixing terms and a bag with mesons and mixing terms.}
\label{fig4}
\end{figure}

In Fig. \ref{fig4}
we show the EMC and earlier SLAC data for $x g_1^p (x)$ together
with the bare bag, the bag plus mesons but without the
$N$-$\Delta$ mixing terms and the bag plus mesons plus
mixing terms.
We stress that the parameters for
the bare bag differ significantly from those used in earlier
calculations \cite{andreas}.
The meson cloud lowers the bag model prediction over the entire
range of $x$, in accordance with earlier estimates for the
Ellis-Jaffe sum rule in the bag \cite{tony4}. This is because some
of the spin of the
nucleon is carried as angular momentum by the mesons.
The actual value of the calculated Ellis-Jaffe
sum rule at $10\; GeV^2$drops from 0.209 in the bare bag to $\sim 0.173$ in the
bag
plus mesons for case (a) and to $\sim 0.169$ for case (b).
The actual value of the sum rule in the bag model plus mesons, calculated using
Eq. (\ref{f16}) as it stands, is $\sim 0.171$, supporting our claim
that the full NLO prediction for the $x$ dependence of $g_{1p}(x)$
in the present model is somewhere between the results for case (a) and (b)
of Fig. \ref{fig4}.
The inclusion of vector mesons
is expected to reduce the value of the sum rule somewhat more.

\begin{figure}[h]
\graphics{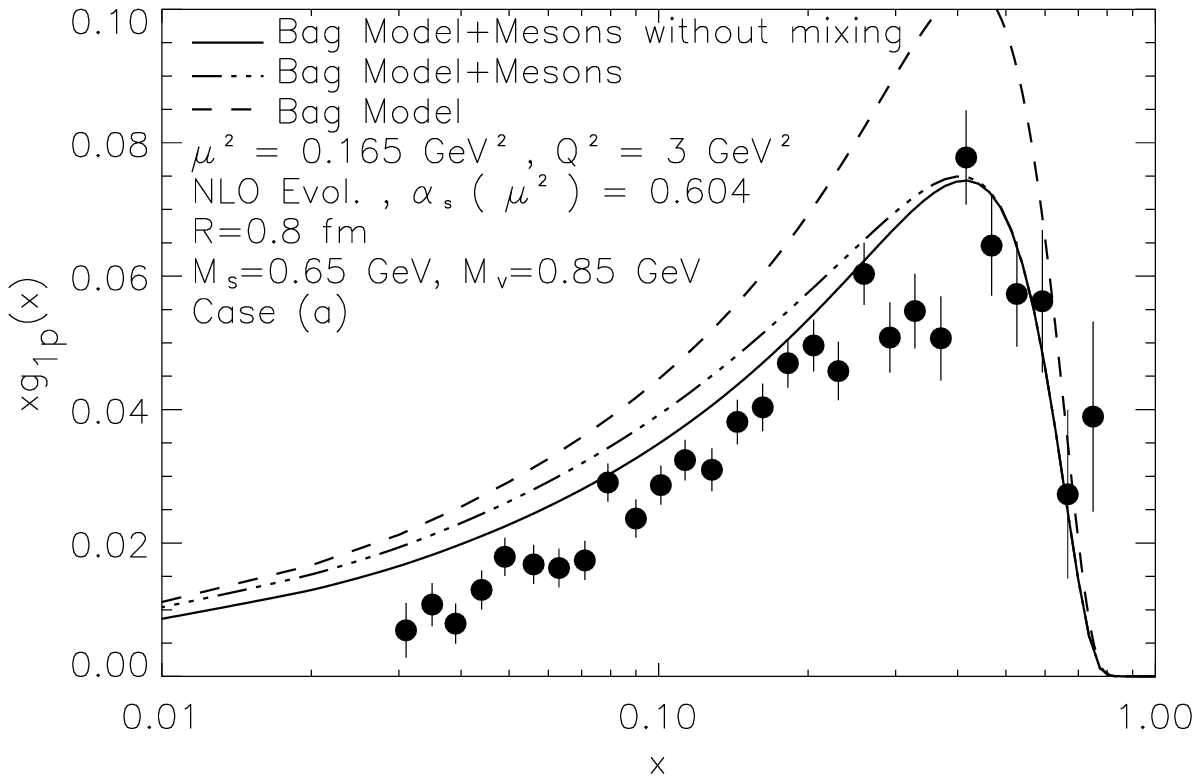}{17}{10}{0}{-10}{1}
\graphics{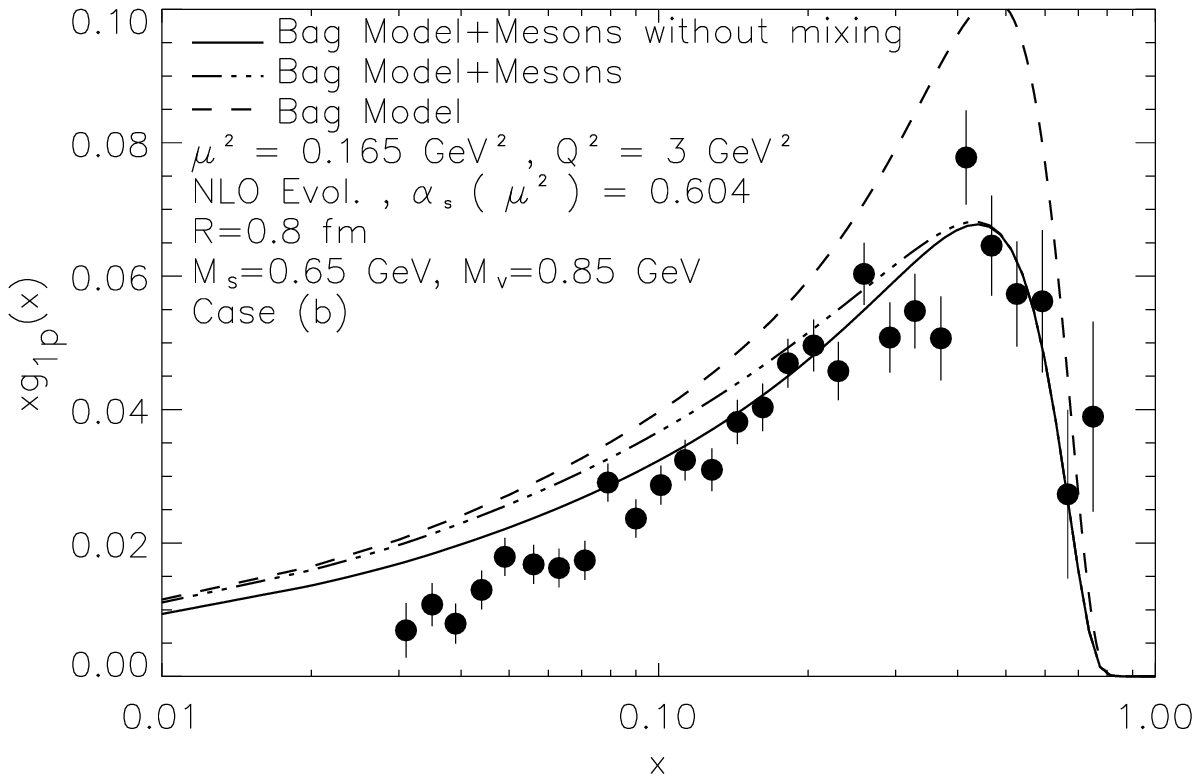}{17}{10}{0}{-10}{1}
\caption{Polarized quark distribution of the proton as
measured by the SLAC-E143 experiment \protect\cite{slac143} against theoretical
predictions for a bare bag, a bag with mesons without
mixing terms and a bag with mesons and mixing terms.}
\label{fig5}
\end{figure}

The $x$ dependence for the model, compared with the SLAC E143
data \cite{slac143} for $g_1^p (x)$, is shown in Fig. \ref{fig5},
and is quite
impressive. This data was taken at an average $3 \; GeV^2$ and has
smaller error bars than earlier experiments. Comparison with this
set is also a good test of our model once we have to move to a different
$Q^2$. The resulting agreement between theoretical and experimental
values for $g_1 (x)$ is inspiring and provides some confidence
in the model. At this value of $Q^2$ the calculated
value for the Ellis-Jaffe sum rule is $\sim 0.171$ for case(a) and
$\sim 0.166$ for case (b).

These results shed light on how the
spin in the physical proton is shared, suggesting strongly that
mesons are responsible for part of the dilution of the spin.
In general, we can say that the agreement between the
data and the theoretical calculation is very
impressive and that
further corrections might well bring the entire curve
within experimental errors. For instance, we know that in the
axial gauge the axial anomaly is related to the polarized
gluon distribution \cite{bass}. In this context, we think that
a reasonable polarized gluon distribution \cite{stirling} could
bring the curve down in the region $x < 0.4$.

We would also like to call attention to the role of the
$N$-$\Delta$ mixing term. As said before, it tends to increase
the Ellis-Jaffe sum rule. In fact, if the mixing terms were absent,
the new value of the sum rule at $Q^2 = 10 \; GeV^2$ would be
$\sim 0.159$ for case (a) and
$\sim 0.155$ for case (b). For the $Q^2$ compatible with the SLAC E143 data,
these values would be
reduced to $\sim 0.158$ for case (a) and to $\sim 0.152$ for case(b).
Figures 2 and 3 tell us
that the rise of $g_1^p (x)$ due to the mixing terms is confined to the region
$x\leq 0.3$. This is because for the mixing term,
the mass of the intermediate state is always $M_v$, and then
the contribution is isolated at smaller $x$ when compared with the other
contributions
(similar to the down quark distribution in the bag \cite{andreas}).

\begin{figure}[h]
\vspace{5cm}
\graphics{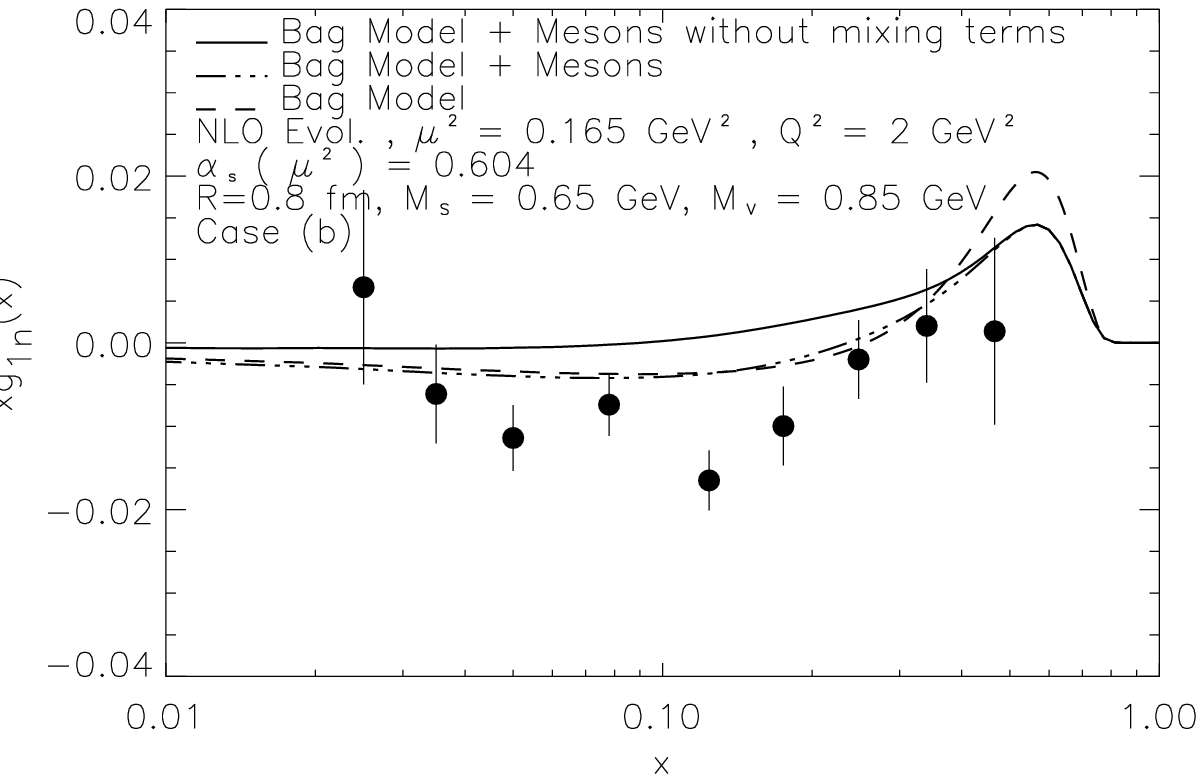}{17}{10}{0}{-10}{1}
\caption{Polarized quark distribution of the neutron as
measured by the SLAC-E142 experiment \protect\cite{slac142} against theoretical
predictions for a bare bag, a bag with mesons without
mixing terms and a bag with mesons and mixing terms.}
\label{fig8}
\end{figure}

The most interesting effect associated with the mixing terms can be observed
in Fig. \ref{fig8},
where $g_1^n (x)$ is shown. Although in $g_1^p (x)$
the effect of the mixing terms in the $x$ distribution is not
too dramatic because one is adding a small number to a large number,
in $g_1^n (x)$ the effect is relatively large because one is
subtracting from numbers near to zero. In fact, for $g_1^n (x)$ the
mixing terms are essential to give to the theoretical
curve the shape of the experimental data as measured by the SLAC E142
\cite{slac142} experiment. The first moment of the calculated
polarized distribution of the neutron turns out to be $\sim 0.004$
for case (a) and $\sim -0.003$ for case (b).
It is also worth noticing that there is a consistency
between the calculation of $g_{1p} (x)$ and $g_{1n} (x)$ --
the same calculation that fits the unpolarized data also makes a good
prediction for both $g_{1p} (x)$ and $g_{1n} (x)$.
\newline
\newline

We would like to thank S. Bass, W. Melnitchouk, G. Piller, A. W. Schreiber
and J. Speth for valuable discussions during the development of this work.
One of the authors (H.H.) would like to thank the German Academic Exchange
Service (DAAD) for financial support for his visit to Adelaide. This work
was supported by the Australian Research Council, by the DAAD and by
CAPES (Brazil).

\appendix
\makeatletter                    
\@addtoreset{equation}{section}  
\makeatother                     
\renewcommand{\theequation}{\thesection.\arabic{equation}}
\section{Appendix}

\begin{eqnarray}
d_{N\pi}(y)={g^2_{NN\pi}\over 16\pi^2}\int_0^\infty dk^2_\perp
{|G_{N\pi}(y,k^2_\perp|^2\over y^2(1-y)}
{m^2_N(1-y)^2-k^2_\perp\over [m_N^2-M^2_{N\pi}(y,k^2_\perp)]^2},
\end{eqnarray}

\begin{eqnarray}
d_{\Delta\pi}(y)&=&{g^2_{N\Delta\pi}\over 96\pi^2}
\int_0^\infty dk^2_\perp
{|G_{\Delta\pi}(y,k^2_\perp)|^2\over y^4(1-y)m_\Delta^2} \nonumber \\*
    &\times&  {[(ym_N+m_\Delta)^2+k^2_\perp]
[(y^2m_N^2-m_\Delta^2)^2+8ym_Nm_\Delta k^2_\perp-k^4_\perp]
\over [m_N^2-M^2_{\Delta\pi}(y,k^2_\perp)]^2}.
\end{eqnarray}
and
\begin{eqnarray}
d_{int}(y)&=&{g_{N\Delta\pi} g_{NN\pi}\over 16\sqrt 6\pi^2}
\int_0^\infty dk^2_\perp
{G_{\Delta\pi}(y,k^2_\perp)G_{N\pi}(y,k^2_\perp) \over y^3(1-y)m_\Delta}
  \nonumber \\*
&\times&\left\{\frac{-m_N(1-y)(ym_N+m_\Delta)^2(ym_N-m_\Delta)}
              {[m_N^2-M^2_{\Delta\pi}(y,k^2_\perp)]
              [m_N^2-M^2_{N\pi}(y,k^2_\perp)]}\right. \nonumber \\*
& & \left. + \frac{(2m_\Delta^2+(3y-2)m_Nm_\Delta-ym_N^2)k^2_\perp-k_\perp^4}
              {[m_N^2-M^2_{\Delta\pi}(y,k^2_\perp)]
      [m_N^2-M^2_{N\pi}(y,k^2_\perp)]}\right\}.
\end{eqnarray}

\addcontentsline{toc}{chapter}{\protect\numberline{}{References}}

\end{document}